\documentclass[12pt]{article}
\usepackage{amsmath,amssymb}
\usepackage{bm}
\usepackage{mathrsfs}
\usepackage{fourier}
\allowdisplaybreaks[4]

\begin{document}

\title{Classification of Standard Model Particles in
$E_6$ Orbifold Grand Unified Theories}

\author{
Yoshiharu \textsc{Kawamura}\footnote{E-mail: haru@azusa.shinshu-u.ac.jp}\\
{\it Department of Physics, Shinshu University, }\\
{\it Matsumoto 390-8621, Japan}\\
and \\
Takashi \textsc{Miura}\footnote{E-mail: takashi.miura@people.kobe-u.ac.jp}\\
{\it Department of Physics, Kobe University, }\\
{\it Kobe 657-8501, Japan}\\
}

\date{
January 31, 2013}

\maketitle
\begin{abstract}
We classify the standard model fermions, which originate from
bulk fields of the $\bf{27}$ or $\overline{\bf{27}}$ representation after orbifold breaking,
in $E_6$ grand unified theories on 5 or 6-dimensional space-time,
under the condition that $q$, $e^c$ and $u^c$ survive as zero modes
for each ${\bf 27}$ or $\overline{\bf{27}}$.
We study features of supersymmetric $SU(5) \times U(1)_1 \times U(1)_2$ model.
\end{abstract}

{\it Keywords}: Grand unified theory; Orbifold breaking.

\section{Introduction}

{\it Grand unification} is attractive, because it offers a unification of force and a (partial-)unification of 
quarks and leptons in each family.\cite{G&G,Langacker,Ross}
In the grand unification based on the $E_6$ gauge group,
$E_6$ has the standard model (SM) gauge group 
$G_{\tiny{\mbox{SM}}} = SU(3)_C \times SU(2)_L \times U(1)_Y$ as a subgroup,
the left-handed multiplet of the $\bf{27}$ representation includes the SM matters in each family,
and the SM Higgs particle can be in a member of bosonic component in the multiplet of $\bf{27}$.\cite{EGUT}

However, lot of extra particles exist in multiplets of $E_6$, e.g., 
66 extra ones in the gauge boson multiplet of $\bf{78}$ and
11 extra ones in each matter multiplet of $\bf{27}$.
In most cases, the unwelcome particles are expected to be heavy 
after the breakdown of $E_6$ into $G_{\tiny{\mbox{SM}}}$
and decouple in the low-energy physics, 
because they are not chiral under $G_{\tiny{\mbox{SM}}}$.
It can be backed by the survival hypothesis.\cite{G}
In this case, it depends on models which particles survive in the SM.

In 4-dimensional models, a Weyl fermion of $\bf{27}$ contains two sets of 
the charge conjugated state ($d^c$) of right-handed down type quark, 
the charged lepton doublet ($l$) and the neutrino singlet ($\nu$) in the SM language, 
and one of them or a linear combination of them would be the SM one.

By the extension of models on a higher-dimensional space-time 
including orbifolds as an extra space,\footnote{
Models based on orbifold were initially utilized on the construction of 4-dimensional
string models.\cite{Orbifold1,Orbifold2,C&K}
Higher-dimensional grand unified theory on orbifold has been proposed 
with several attractive features.\cite{K,H&N1,H&N2}
Higher-dimensional $E_6$ grand unified theories on orbifold have been studied
from several aspects.\cite{HJL&L,H&S,H&R,FN&W,BK&R}
}$^,$\footnote{
The constructions of low-energy theory have been make through
dimensional reduction over coset space.\cite{K&Z,I&Z,IO&Z}
} 
two additional features are provided.
One is that right-handed Weyl fermions can also appear after compactification, 
because a fermion on the higher-dimensional space-time contains 
both left-handed and right-handed components in terms of 4-dimensional Weyl fermion.
Hence a fermion of $\bf{27}$ can contain mirror particles for $d^c$ and $l$.
Some of them (which would be right-handed components) can be regarded as 
$d^c$ and $l$ (which would be left-handed ones) after the charge conjugation
has been carried out.
In a similar way, $\overline{\bf{27}}$ can be also useful on the model-building,
because SM fermions can originate from it.
The other is that 4-dimensional fields are regarded as zero modes 
after imposing orbifold boundary condition.
In this process, the survival hypothesis does not necessarily work, 
and it depends on boundary conditions which particles survive in the low energy theory.

Considering the above features, it is interesting to study 
under what type of boundary conditions SM matters survive after orbifolding.
In this paper, we classify the SM particles, which originate from
bulk fields of $\bf{27}$ or $\overline{\bf{27}}$ after orbifold breaking,
in $E_6$ grand unified theories on 5 or 6-dimensional space-time,
under the condition that the quark doublet ($q$), 
the charge conjugated state ($e^c$) of right-handed electron type lepton
and the charge conjugated state ($u^c$) of right-handed up type quark survive as zero modes
for each $\bf{27}$ or $\overline{\bf{27}}$.
The analysis is carried out based on the subgroup $SU(3) \times SU(3) \times SU(3)$ of $E_6$ 
and the diagonal embedding of $Z_M$ orbifolding ($M=2, 3, 4, 6$).

The outline of our paper is as follows.
In Sec. 2, we explain features of $E_6$ grand unification
and $Z_N$ orbifold breaking.
We classify the SM matters, which originate from
bulk fields of $\bf{27}$ or $\overline{\bf{27}}$ 
by orbifolding in Sec. 3.
In Sec. 4, we study features of effective supersymmetric $SU(5) \times U(1)_1 \times U(1)_2$ model
after orbifolding.
In the last section, conclusions and a discussion are presented.

\section{$E_6$ grand unification and $Z_N$ orbifold breaking}

\subsection{$E_6$ grand unification}

$E_6$ has three types of maximal subgroup such as $SO(10) \times U(1)$,
$SU(6) \times SU(2)$ and $SU(3) \times SU(3) \times SU(3)$.\cite{Slansky}
We consider the subgroup $G_{\tiny{\mbox{Tri}}} = SU(3)_C \times SU(3)_L \times SU(3)_X$.
Here, the first $SU(3)$ is identified with $SU(3)_C$,
the second one contains $SU(2)_L$ as a subgroup,
and the third one is denoted by $SU(3)_X$.
Under $G_{\tiny{\mbox{Tri}}}$ and 
its subgroup $G_{\tiny{\mbox{32111}}} = SU(3)_C \times SU(2)_L \times U(1)_{L8} \times U(1)_{X3} \times U(1)_{X8}$, 
The $\bf{27}$ and $\overline{\bf{27}}$ representations are 
decomposed into a sum of multiplets such that
\begin{eqnarray}
&~& {\bf{27}} = ({\bf{3}}, {\bf{3}}, {\bf{1}}) + ({\bf{1}}, \overline{\bf{3}}, {\bf{3}})
 + (\overline{\bf{3}}, {\bf{1}}, \overline{\bf{3}})~,
\label{27-Tri}\\
&~& ~~~~ ({\bf{3}}, {\bf{3}}, {\bf{1}}) = ({\bf{3}}, {\bf{2}})_{1/3,0,0} + ({\bf{3}}, {\bf{1}})_{-2/3,0,0}~,
\nonumber \\
&~& ~~~~ ({\bf{1}}, \overline{\bf{3}}, {\bf{3}}) 
= ({\bf{1}}, {\bf{2}})_{-1/3, 1, 1/3} + ({\bf{1}}, {\bf{2}})_{-1/3, -1, 1/3} + ({\bf{1}}, {\bf{2}})_{-1/3, 0, -2/3}
\nonumber \\
&~& ~~~~~~~~~~~~~~~~~~~~~~ 
+ ({\bf{1}}, {\bf{1}})_{2/3, 1, 1/3} + ({\bf{1}}, {\bf{1}})_{2/3, -1, 1/3} 
+ ({\bf{1}}, {\bf{1}})_{2/3, 0, -2/3}~,
\nonumber \\
&~& ~~~~ (\overline{\bf{3}}, {\bf{1}}, \overline{\bf{3}})
= (\overline{\bf{3}}, {\bf{1}})_{0,-1,-1/3} + (\overline{\bf{3}}, {\bf{1}})_{0,1,-1/3}
+ (\overline{\bf{3}}, {\bf{1}})_{0,0,2/3}
\label{27-SM}
\end{eqnarray}
and
\begin{eqnarray}
&~& \overline{\bf{27}} = (\overline{\bf{3}}, \overline{\bf{3}}, {\bf{1}}) + ({\bf{1}}, {\bf{3}}, \overline{\bf{3}})
 + ({\bf{3}}, {\bf{1}}, {\bf{3}})~,
\label{bar27-Tri}\\
&~& ~~~~ (\overline{\bf{3}}, \overline{\bf{3}}, {\bf{1}}) 
= (\overline{\bf{3}}, {\bf{2}})_{-1/3,0,0} + (\overline{\bf{3}}, {\bf{1}})_{2/3,0,0}~,
\nonumber \\
&~& ~~~~ ({\bf{1}}, {\bf{3}}, \overline{\bf{3}}) 
= ({\bf{1}}, {\bf{2}})_{1/3, -1, -1/3} + ({\bf{1}}, {\bf{2}})_{1/3, 1, -1/3} + ({\bf{1}}, {\bf{2}})_{1/3, 0, 2/3}
\nonumber \\
&~& ~~~~~~~~~~~~~~~~~~~~~~ 
+({\bf{1}}, {\bf{1}})_{-2/3, -1, -1/3} + ({\bf{1}}, {\bf{1}})_{-2/3, 1, -1/3} 
+ ({\bf{1}}, {\bf{1}})_{-2/3, 0, 2/3}~,
\nonumber \\
&~& ~~~~ ({\bf{3}}, {\bf{1}}, {\bf{3}})
= ({\bf{3}}, {\bf{1}})_{0,1,1/3} + ({\bf{3}}, {\bf{1}})_{0,-1,1/3} + ({\bf{3}}, {\bf{1}})_{0,0,-2/3}~,
\label{bar27-SM}
\end{eqnarray}
where representations of $G_{\tiny{\mbox{Tri}}}$ and $G_{\tiny{\mbox{32111}}}$
are denoted as $(SU(3)_C, SU(3)_L, SU(3)_X)$ in (\ref{27-Tri}) and (\ref{bar27-Tri})
and as $(SU(3)_C, SU(2)_L)_{Q_{L8}, Q_{X3}, Q_{X8}}$ in (\ref{27-SM}) and (\ref{bar27-SM}), respectively.
Here, $Q_{L8}$, $Q_{X3}$, and $Q_{X8}$ are $U(1)$ charges relating $U(1)_{L8}$, $U(1)_{X3}$ and $U(1)_{X8}$, respectively.

If we require that one family of SM matters should be included in a $\bf{27}$,
there exist three types of definition for the hypercharge $Y$ in the SM such that
\begin{eqnarray}
&~& Y = \frac{1}{2}\left(Q_{L8} + Q_{X3} + Q_{X8}\right) \equiv Y_{(1)}~,~~
Y = \frac{1}{2}\left(Q_{L8} - Q_{X3} + Q_{X8}\right) \equiv Y_{(2)}~,~~
\nonumber \\
&~& Y = \frac{Q_{L8}}{2} - Q_{X8} \equiv Y_{(3)}~.
\label{Y}
\end{eqnarray}
The representations are given based on $Y_{(1)}$ in Table \ref{t2}.
In Table \ref{t2}, fermions denoted by $q$, $u^c$, $d^c_a$, $l_a$, $e^c$ and $\nu^c_a$ $(a = 1, 2)$
have the SM gauge quantum numbers such as $({\bf{3}}, {\bf{2}}, 1/6)$,
$(\overline{\bf{3}}, {\bf{1}}, -2/3)$, $(\overline{\bf{3}}, {\bf{1}}, 1/3)$, $({\bf{1}}, {\bf{2}}, -1/2)$,
$({\bf{1}}, {\bf{1}}, 1)$ and $({\bf{1}}, {\bf{1}}, 0)$,
and some of them are expected to become members in one family.
Note that $q$, $e^c$ and $u^c$ belong to $({\bf{3}}, {\bf{3}}, {\bf{1}})$, $({\bf{1}}, \overline{\bf{3}}, {\bf{3}})$
and $(\overline{\bf{3}}, {\bf{1}}, \overline{\bf{3}})$, respectively.
As mentioned in the introduction, $\bf{27}$ contains two kinds of $d^c$, $l$ and $\nu^c$,
and the label $a$ is attached to distinguish them.
Their mirror particles are denoted by $Q^c$, $U$, $D_a$, $L^c_a$, $E$ and $N_a$
whose SM gauge quantum numbers are given by $(\overline{\bf{3}}, {\bf{2}}, -1/6)$,
$({\bf{3}}, {\bf{1}}, 2/3)$, $({\bf{3}}, {\bf{1}}, -1/3)$, $({\bf{1}}, {\bf{2}}, 1/2)$,
$({\bf{1}}, {\bf{1}}, -1)$ and $({\bf{1}}, {\bf{1}}, 0)$, respectively.
Note that both $\nu^c_a$ and $N_a$ are candidates of singlet neutrinos.

For the definition $Y_{(2)}$, $L^c$, $e^c$ and $u^c$ 
should be exchanged into $l_1$, $\nu^c_1$ and $d^c_1$, respectively.
For the definition $Y_{(3)}$, $L^c$, $e^c$ and $u^c$ 
should be exchanged into $l_2$, $\nu^c_2$ and $d^c_2$, respectively.\footnote{
These features are understood from the existence of $E$-symmetry group $SU(2)_E$ defined in \cite{B&K}.
}

For the gauge bosons, the $\bf{78}$ repesentation is decomposed into a sum of multiplets such that
\begin{eqnarray}
&~& {\bf{78}} = ({\bf{8}}, {\bf{1}}, {\bf{1}}) + ({\bf{1}}, {\bf{8}}, {\bf{1}})
+ ({\bf{1}}, {\bf{1}}, {\bf{8}})
 + ({\bf{3}}, \overline{\bf{3}}, \overline{\bf{3}})
 + (\overline{\bf{3}}, {\bf{3}}, {\bf{3}})~,
\label{78-Tri}\\
&~& ~~~~ ({\bf{8}}, {\bf{1}}, {\bf{1}}) = ({\bf{8}}, {\bf{1}})_{0,0,0}~,
\nonumber \\
&~& ~~~~ ({\bf{1}}, {\bf{8}}, {\bf{1}}) = ({\bf{1}}, {\bf{3}})_{0,0,0}
+ ({\bf{1}}, {\bf{2}})_{1,0,0} + ({\bf{1}}, {\bf{2}})_{-1,0,0}
+ ({\bf{1}}, {\bf{1}})_{0,0,0}~,
\nonumber \\
&~& ~~~~ ({\bf{1}}, {\bf{1}}, {\bf{8}}) = ({\bf{1}}, {\bf{1}})_{0,2,0}
+ ({\bf{1}}, {\bf{1}})_{0,0,0} + ({\bf{1}}, {\bf{1}})_{0,-2,0}
+ ({\bf{1}}, {\bf{1}})_{0,1,1} 
\nonumber \\
&~& ~~~~~~~~~~~~~~~~~~~~~~ + ({\bf{1}}, {\bf{1}})_{0,-1,1}
+ ({\bf{1}}, {\bf{1}})_{0,1,-1} + ({\bf{1}}, {\bf{1}})_{0,-1,-1}
+ ({\bf{1}}, {\bf{1}})_{0,0,0}~,
\nonumber \\
&~& ~~~~ ({\bf{3}}, \overline{\bf{3}}, \overline{\bf{3}})
= ({\bf{3}}, {\bf{2}})_{-1/3, -1, -1/3} + ({\bf{3}}, {\bf{2}})_{-1/3, 1, -1/3} + ({\bf{3}}, {\bf{2}})_{-1/3, 0, 2/3}
\nonumber \\
&~& ~~~~~~~~~~~~~~~~~~~~~~ 
+ ({\bf{3}}, {\bf{1}})_{2/3, -1, -1/3} + ({\bf{3}}, {\bf{1}})_{2/3, 1, -1/3} 
+ ({\bf{3}}, {\bf{1}})_{2/3, 0, 2/3}~,
\nonumber \\
&~& ~~~~ (\overline{\bf{3}}, {\bf{3}}, {\bf{3}})
= (\overline{\bf{3}}, {\bf{2}})_{1/3,1,1/3} + (\overline{\bf{3}}, {\bf{2}})_{1/3,-1,1/3}
+ (\overline{\bf{3}}, {\bf{2}})_{1/3,0,-2/3}
\nonumber \\
&~& ~~~~~~~~~~~~~~~~~~~~~~ 
+ (\overline{\bf{3}}, {\bf{1}})_{-2/3,1,1/3} + (\overline{\bf{3}}, {\bf{1}})_{-2/3,-1,1/3}
+ (\overline{\bf{3}}, {\bf{1}})_{-2/3,0,-2/3}~,
\label{78-SM}
\end{eqnarray}
under $G_{\tiny{\mbox{Tri}}}$ and $G_{\tiny{\mbox{32111}}}$, respectively.

\subsection{$Z_N$ orbifold breaking}

First let us consider 1-dimensional orbifold $S^1/Z_2$ as an example.
The $S^1/Z_2$ is obtained by dividing the circle $S^1$ 
(with the identification $y \sim y + 2\pi R$ through the translation $T:y \to y + 2\pi R$) 
by the $Z_2$ transformation $Z_2:y \to -y$ so that the point $y$ is identified with $-y$.
Both end points $y=0$ and $\pi R$ are fixed points under the $Z_2$ transformation.
The operations are also characterized by $Z_2$ and $Z'_2 (= Z_2 T):y \to 2 \pi R - y$ 
in place of $Z_2$ and $T$.

Accompanied by the identification of points on the extra space,
the following boundary conditions for a field $\Phi(x, y)$ can be imposed on
\begin{eqnarray}
\Phi(x, -y) =T_{\Phi}[P_0] \Phi(x,y)~,~~ \Phi(x, 2\pi R-y) =T_{\Phi}[P_1] \Phi(x,y)~,
\label{T[P]}
\end{eqnarray}
where $T_{\Phi}[P_0]$ and $T_{\Phi}[P_1]$ represent appropriate representation matrices
with $P_0$ and $P_1$ standing for the representation matrices of the fundamental representation
for the $Z_2$ and $Z'_2$ transformation, respectively.
The representation matrices satisfy the relations $T_{\Phi}[P_0]^2 =I$ and $T_{\Phi}[P_1]^2 =I$
because of the $Z_2$ symmetry property,
where $I$ is the unit matrix.

The eigenvalues of $T_{\Phi}[P_0]$ and $T_{\Phi}[P_1]$ are interpreted as the $Z_2$ parity for 
the extra space.
The fields with even $Z_2$ parities have zero modes, but those including an odd $Z_2$ parity 
have no zero modes.
Here, zero modes mean 4-dimensional massless fields surviving after compactification.
Massive Kaluza-Klein modes do not appear in our low-energy world,
because they have heavy masses of $O(1/R)$, with the same magnitude 
as the unification scale.
{\it Unless all components of non-singlet field have a common $Z_2$ parity, 
a symmetry reduction occurs upon compactification because zero modes are absent
in fields with an odd parity.}
This type of symmetry breaking mechanism is called $\lq\lq$orbifold breaking mechanism".\footnote{
The $Z_2$ orbifolding was used in superstring theory\cite{A} and heterotic $M$-theory.\cite{H&W1,H&W2}
In field theoretical models, it was applied to the reduction of global SUSY,\cite{M&P,P&Q} which is
an orbifold version of Scherk-Schwarz mechanism,\cite{S&S,S&S2} and then
to the reduction of gauge symmetry.\cite{K1}
}
The orbifold breaking on $S^1/Z_2$ is characterized by $P_0$ and $P_1$.

Next we consider 2-dimensional orbifold $S^1/Z_2 \times S^1/Z_2$ 
and $T^2/Z_N$ $(N=2, 3, 4, 6)$ in order.

The orbifold breaking on $S^1/Z_2 \times S^1/Z_2$ is characterized by 
two pairs of $Z_2$ transformation matrices denoted by $(P_{10}, P_{11})$ and $(P_{20}, P_{21})$.\cite{K&M}

Let $z$ be the complex coordinate of $T^2/Z_N$.
Here, $T^2$ is constructed using a 2-dimensional lattice.
On $T^2$, the points $z + e_1$ and $z + e_2$ are identified with
the point $z$ where $e_1$ and $e_2$ are basis vectors.
The orbifold $T^2/Z_N$ is obtained by dividing $T^2$ by the $Z_N$ transformation
$Z_N:z \to \rho z$ $(\rho^N = 1)$ so that the point $z$ is identified with $\rho z$.

For $T^2/Z_2$, basis vectors are given by $e_1 =1$ and $e_2 =i$,
and the orbifold breaking is featured by $P_0$, $P_1$ and $P_2$,
which are related to the $Z_2$ transformations $z \to -z$, $z \to e_1-z$
and $z \to e_2-z$, respectively.
Basis vectors, representation matrices and their transformation property of $T^2/Z_N$
are summarized in Table 1.\cite{K&M,K&M1}\footnote{
Though the number of independent representation matrices for $T^2/Z_6$ is stated to be three in \cite{K&M},
it should be two because other operations are generated using $s_0:z \to e^{\pi i/3} z$ and $r_1:z \to e_1-z$.
For example, $t_1:z \to z+e_1$ and $t_2:z \to z+e_2$ are generated as $t_1= r_1 (s_0)^3$
and $t_2 = (s_0)^2 r_1 (s_0)^4 r_1$, respectively.
}
Note that there is a choice of representation matrices and
$P_1$ for the $Z_2$ transformation $z \to e_1-z$ is also used in $T^2/Z_4$ and $T^2/Z_6$.
Fields possess discrete charges relating eigenvalues of representation matrices
for $Z_M$ transformation.
Here, $M=N$ for $N=2,3$ and $M=N,2$ for $N=4,6$.
\begin{table}[htb]
\caption{The characters of $T^2/Z_N$.}
\label{t1}
\begin{center}
\begin{tabular}{c|c|c|c} \hline
$N$ & {\it Basis vectors} & {\it Rep. matrices} & {\it Transformation property} \\ \hline\hline
$2$ & $1, i$ & $P_0,~ P_1,~ P_2$ & $z \to -z,~ z \to e_1-z,~ z \to e_2-z$ \\ \hline
$3$ & $1, e^{2\pi i/3}$ & $\Theta_0,~ \Theta_1$ & $z \to e^{2\pi i/3} z,~ z \to e^{2\pi i/3} z + e_1$ \\ \hline
$4$ & $1, i$ & $Q_0,~ P_1$ & $z \to iz,~ z \to e_1-z$ \\ \hline
$6$ & $1, (-3+i\sqrt{3})/2$ & $\Xi_0,~ P_{1}$ & $z \to e^{\pi i/3} z,~ z \to e_1-z$ \\ \hline
\end{tabular}
\end{center}
\end{table}

\subsection{Elements for $Z_M$ transformation}

We explain the assignment of discrete charge or element for $Z_M$ transformation
using the breakdown of $SU(3)$ into its subgroups as an example.

In the case with the representation matrix
\begin{eqnarray}
R_a = \mbox{diag}(\rho_1, \rho_1, \rho_2)~,~~ \rho_1 \ne \rho_2~,
\label{Ra}
\end{eqnarray}
$SU(3)$ is broken down to $SU(2) \times U(1)$.
Here, $\rho_i$s are elements of $Z_M$, i.e., $\rho_i^M = 1$
and we refer them to $Z_M$ elements.\footnote{
The $Z_M$ elements are given by $e^{2\pi i n/M}$ $(n=0, 1, \cdots, M-1)$,
and the number $n/M$ is the charge for $Z_M$ transformation, which is usually called $Z_M$ charge.}
Then the fundamental representation ${\bf 3}$ is decomposed into a sum of multiplets such that
\begin{eqnarray}
{\bf{3}} = {\bf{2}}_{1/3}\left(\eta \rho_1\right) 
~+~ {\bf{1}}_{-2/3}\left(\eta \rho_2\right)~,
\label{3rep}
\end{eqnarray}
where numbers listed in bold font in the right hand side stand for representations of $SU(2)$, 
numbers indicated by a subscript are $U(1)$ charge
and symbols in the bracket are $Z_M$ elements ($\eta$ is the intrinsic $Z_M$ element of $\bf{3}$).
The conjugate representation $\overline{\bf{3}}$ is equivalent to the antisymmetric part of
the product $\bf{3} \times \bf{3}$, and hence $\overline{\bf{3}}$ is decomposed into a sum of multiplets such that
\begin{eqnarray}
&~& \overline{\bf{3}} = \left[({\bf{2}}_{1/3}+{\bf{1}}_{-2/3}) \times ({\bf{2}}_{1/3}+{\bf{1}}_{-2/3})\right]_{A}
\nonumber \\
&~& ~~ = \left[({\bf{2}}_{1/3} \times {\bf{1}}_{-2/3}) + ({\bf{1}}_{-2/3} \times {\bf{2}}_{1/3})\right]_{A}
+ \left[{\bf{2}}_{1/3} \times {\bf{2}}_{1/3}\right]_{A} 
\nonumber \\
&~& ~~ = {\bf{2}}_{-1/3}\left(\tilde{\eta} \rho_1 \rho_2\right) 
~+~ {\bf{1}}_{2/3}\left(\tilde{\eta} (\rho_1)^2\right)~,
\label{bar3rep}
\end{eqnarray}
where the subscript ($A$) represents antisymmetric part and
$\tilde{\eta}$ is the intrinsic $Z_M$ element of $\overline{\bf{3}}$.
Using $(\rho_1)^2 \rho_2 = \alpha$, (\ref{bar3rep}) is rewritten by
\begin{eqnarray}
\overline{\bf{3}} = {\bf{2}}_{-1/3}\left(\tilde{\eta}' \overline{\rho}_1\right) 
~+~ {\bf{1}}_{2/3}\left(\tilde{\eta}' \overline{\rho}_2\right)~,
\label{bar3rep2}
\end{eqnarray}
where $\overline{\rho}_i$ ($i=1, 2$) is the complex conjugation of $\rho_i$
and $\tilde{\eta}' = \tilde{\eta} \alpha$.

In a similar way, for the representation matrix
\begin{eqnarray}
R_b = \mbox{diag}(\rho_1, \rho_2, \rho_3)~,~~ \rho_1 \ne \rho_2~,~~ \rho_1 \ne \rho_3~,~~ \rho_2 \ne \rho_3~,
\label{Rb}
\end{eqnarray}
$SU(3)$ is broken down to $U(1) \times U(1)$, and
then the ${\bf 3}$ is decomposed into a sum of multiplets such that
\begin{eqnarray}
{\bf{3}} = \left(1,1/3;~ \eta \rho_1\right) 
~+~ \left(-1,1/3;~ \eta \rho_2\right) 
~+~ \left(0,-2/3;~ \eta \rho_3\right)~,
\label{3rep-b}
\end{eqnarray}
where numbers in the right hand side stand for representations of $U(1)$ charges 
and symbols are $Z_M$ elements including the intrinsic $Z_M$ element $\eta$.
The $\overline{\bf{3}}$ with the intrinsic $Z_M$ element $\tilde{\eta}$ 
is decomposed into a sum of multiplets such that
\begin{eqnarray}
&~& \overline{\bf{3}} = \left(0,2/3;~ \tilde{\eta} \rho_1 \rho_2\right) 
~+~ \left(-1,-1/3;~ \tilde{\eta} \rho_2 \rho_3\right) 
~+~ \left(1,-1/3;~ \tilde{\eta} \rho_1 \rho_3\right)~.
\label{bar3rep-b}
\end{eqnarray}
Using $\rho_1 \rho_2 \rho_3 = \beta$, (\ref{bar3rep-b}) is rewritten by
\begin{eqnarray}
\overline{\bf{3}} = \left(0,2/3;~ \tilde{\eta}'' \overline{\rho}_3\right) 
~+~ \left(-1,-1/3;~ \tilde{\eta}'' \overline{\rho}_1\right) 
~+~ \left(1,-1/3;~ \tilde{\eta}'' \overline{\rho}_2\right)~,
\label{bar3rep2-b}
\end{eqnarray}
where $\overline{\rho}_j$ ($j=1, 2, 3$) is the complex conjugation of $\rho_j$
and $\tilde{\eta}'' = \tilde{\eta} \beta$.

Finally, we explain a fermion on 6-dimensional spacetime.
We use the metric $\eta_{MN} = \mbox{diag}(1, -1, -1, -1, -1, -1)$ $(M,N=0,1,2,3,5,6)$
and the following representation for 6-dimensional gamma matrices:
\begin{eqnarray}
\Gamma^{\mu} = \gamma^{\mu} \otimes \sigma_3~, ~~
\Gamma^{5} = I_{4 \times 4}  \otimes i \sigma_1~, ~~ 
\Gamma^{6} = I_{4 \times 4}  \otimes i \sigma_2~,
\label{6Dgamma}
\end{eqnarray}
where $\mu=0, 1, 2, 3$ and $I_{4 \times 4}$ is the $4 \times 4$ unit matrix.
The $\Gamma^M$s satisfy the Clifford algebra $\{\Gamma^M, \Gamma^N\}=2 \eta^{MN}$
where $\eta^{MN}$ is the inverse of $\eta_{MN}$.
The chirality operator $\Gamma_7$ for 6-dimensional fermion $\Psi$ is defined as
\begin{eqnarray}
\Gamma_7 \equiv \Gamma^0 \Gamma^1 \Gamma^2 \Gamma^3 \Gamma^5 \Gamma^6 = -\gamma_{5} \otimes \sigma_3~,
\label{6Dgamma7}
\end{eqnarray}
where $\gamma_5$ is the chirality operator for 4--dimensional fermion.
The fermion $(\Psi_+)$ with positive chirality and the fermion $(\Psi_-)$ with negative chirality
are given by
\begin{eqnarray}
&~& \Psi_+ = \frac{1+\Gamma_7}{2} \Psi 
= \left(
\begin{array}{cc}
\frac{1-\gamma_5}{2} & 0 \\
0 & \frac{1+\gamma_5}{2} 
\end{array}
\right)
\left( 
\begin{array}{c}
\Psi^1 \\
\Psi^2 
\end{array}
\right)
= \left( 
\begin{array}{c}
\Psi^1_L \\
\Psi^2_R 
\end{array}
\right)~,
\label{Psi+}\\
&~& \Psi_- = \frac{1-\Gamma_7}{2} \Psi
= \left(
\begin{array}{cc}
\frac{1+\gamma_5}{2} & 0 \\
0 & \frac{1-\gamma_5}{2} 
\end{array}
\right)
\left( 
\begin{array}{c}
\Psi^1 \\
\Psi^2 
\end{array}
\right)
= \left( 
\begin{array}{c}
\Psi^1_R \\
\Psi^2_L 
\end{array}
\right)~,
\label{Psi-}
\end{eqnarray}
respectively.

In terms of 4-dimensional Weyl fermions $\Psi^1_L$, $\Psi^2_R$, $\Psi^1_R$ and $\Psi^2_L$,
the kinetic terms for $\Psi_+$ and $\Psi_-$ are rewritten as
\begin{eqnarray}
\hspace{-1.5cm} &~& i \overline{\Psi}_+ \Gamma^M D_{M} \Psi_+
= i \overline{\Psi}_+ \Gamma^{\mu} D_{\mu} \Psi_+ 
+ i \overline{\Psi}_+ \Gamma^z D_{z} \Psi_+ + i \overline{\Psi}_+ \Gamma^{\overline{z}} D_{\overline{z}} \Psi_+
\nonumber \\
\hspace{-1.5cm} &~& ~~~~~~~~~~~~~~~~~~~~~~~~~~~~~ 
= i \overline{\Psi}^1_L \gamma^{\mu} D_{\mu} \Psi^1_L
+ i \overline{\Psi}^2_R \gamma^{\mu} D_{\mu} \Psi^2_R
- \overline{\Psi}^1_L D_z \Psi^2_R
+ \overline{\Psi}^2_R D_{\overline{z}} \Psi^1_L ~,
\label{Psi+kinetic}\\
\hspace{-1.5cm} &~& i \overline{\Psi}_- \Gamma^M D_{M} \Psi_-
= i \overline{\Psi}_- \Gamma^{\mu} D_{\mu} \Psi_- 
+ i \overline{\Psi}_- \Gamma^z D_{z} \Psi_- + i \overline{\Psi}_- \Gamma^{\overline{z}} D_{\overline{z}} \Psi_-
\nonumber \\
\hspace{-1.5cm} &~& ~~~~~~~~~~~~~~~~~~~~~~~~~~~~~ 
= i \overline{\Psi}^1_R \gamma^{\mu} D_{\mu} \Psi^1_R
+ i \overline{\Psi}^2_L \gamma^{\mu} D_{\mu} \Psi^2_L
- \overline{\Psi}^1_R D_z \Psi^2_L
+ \overline{\Psi}^2_L D_{\overline{z}} \Psi^1_R ~,
\label{Psi-kinetic}
\end{eqnarray}
where $\overline{\Psi}_+$, $\overline{\Psi}_-$, $\Gamma^{z}$ and $\Gamma^{\overline{z}}$ are defined by
\begin{eqnarray}
&~& \overline{\Psi}_+ \equiv \Psi_+^{\dagger} \Gamma^{0}
= \left(\Psi^{1\dagger}_L \gamma^0, - \Psi^{2\dagger}_R \gamma^0 \right) 
= \left(\overline{\Psi}^{1}_L, -\overline{\Psi}^{2}_R \right)~,
\nonumber \\
&~& \overline{\Psi}_- \equiv \Psi_-^{\dagger} \Gamma^{0}
= \left(\Psi^{1\dagger}_R \gamma^0, - \Psi^{2\dagger}_L \gamma^0 \right) 
= \left(\overline{\Psi}^{1}_R, -\overline{\Psi}^{2}_L \right)~,
\label{6DoverlinePsi} \\ 
&~& \Gamma^{z} \equiv \frac{1}{2}\left(\Gamma^{5} + i \Gamma^{6}\right) = i I_{4 \times 4}  \otimes \sigma_{+}~, ~~
\Gamma^{\overline{z}} \equiv \frac{1}{2}\left(\Gamma^{5} - i \Gamma^{6}\right) = i I_{4 \times 4}  \otimes \sigma_{-}
\label{6Dgamma-z}
\end{eqnarray}
and $z \equiv x^5 + i x^6$ and $\overline{z} \equiv x^5 - i x^6$.
The Kaluza-Klein masses are generated from the terms including $D_z$ and $D_{\overline{z}}$
upon compactification.

There are two choices of assignment for 4-dimensional SM fermions from 6-dimensional Weyl fermion, i.e.,
$\Psi^1_L$ or $\Psi^2_L$.
Let us take $\Psi^1_L$ as the SM fermions.
The $Z_M$ element of $(\Psi^2_{R})^c$ (charge conjugation of $\Psi^2_{R}$) 
is the complex conjugation of $\Psi^2_{R}$,
and that of $\Psi^2_{R}$ is determined by the $Z_M$ invariance of the kinetic term (\ref{Psi+kinetic}) 
and the transformation property of the covariant derivative 
$Z_M : D_z \to \overline{\rho} D_z$ and $D_{\overline{z}} \to \rho D_{\overline{z}}$
with $\overline{\rho} = e^{-2\pi i/M}$ and $\rho = e^{2\pi i/M}$.
{}From these observations, the following relation holds between
the $Z_M$ element of $\Psi^1_L$ and that of its mirror fermion ($(\Psi^2_R)^c$),
\begin{eqnarray}
\mathcal{P}_{(\Psi^2_R)^c} = \overline{\rho} \overline{\mathcal{P}}_{\Psi^1_{L}}~,
\label{rhoPsi2R}
\end{eqnarray}
where $\overline{\mathcal{P}}_{\Psi^1_{L}}$ is the complex conjugation of $\mathcal{P}_{\Psi^1_{L}}$.
If we take $\Psi^2_L$ as the SM fermions, 
the following relation holds between
the $Z_M$ element of $\Psi^2_L$ and that of its mirror fermion ($(\Psi^1_R)^c$),
\begin{eqnarray}
\mathcal{P}_{(\Psi^1_R)^c} = {\rho} \overline{\mathcal{P}}_{\Psi^2_{L}}~,
\label{rhoPsi1R}
\end{eqnarray}
where $\overline{\mathcal{P}}_{\Psi^2_{L}}$ is the complex conjugation of $\mathcal{P}_{\Psi^2_{L}}$.

The above choices of assignment for $\Psi^1_L$ or $\Psi^2_L$ lead to same results
for species with zero modes and unbroken gauge group,
because they are related each other by exchanging between $\rho = e^{2\pi i/M}$ and $\overline{\rho} = e^{-2\pi i/M}$.
The construction of $Z_N$ orbifold does not depend on the choice of $\rho(\ne 1)$. 
Hence we take $\Psi^2_L$ as the SM fermions, in the following.

\section{Classification of SM particles}

\subsection{Assignment of $Z_M$ elements}

The orbifold breaking is characterized by a set of representation matrices.
In most cases, orbifold breaking is analyzed using shift embeddings on roots of $E_6$.\cite{H&R,FN&W,BK&R}
The shift embeddings are useful to classify the breaking pattern of gauge symmetry.
We use the diagonal embedding on the subgroup $G_{\tiny{\mbox{Tri}}} = SU(3)_C \times SU(3)_L \times SU(3)_X$ of $E_6$,
because it has a usability on examining zero modes of matter fields systematically.

Let us take the representation matrix
\begin{eqnarray}
R_M = \mbox{diag}(\rho_1, \rho_1, \rho_1) \times  
\mbox{diag}(\rho_2, \rho_2, \rho_3) \times \mbox{diag}(\rho_4, \rho_5, \rho_6)
\label{R}
\end{eqnarray}
to keep $SU(3)_C$ and $SU(2)_L$ unbroken.
Here, $\rho_i$s are elements of $Z_M$.

The species and $Z_M$ element for matters derived from $\bf{27}$ are assigned in Table \ref{t2}.
\begin{table}[htb]
\caption{The species and $Z_M$ element for matters derived from $\bf{27}$.}
\label{t2}
\begin{center}
\begin{tabular}{l|c|c|c||c|c} \hline
\it{Representations} & $Y_{(1)}$ & $f_{L}$ 
& $\mathcal{P}_{f_{L}}$ &  $(f_{R})^c$ & $\mathcal{P}_{(f_{R})^c}$ \\ \hline\hline
$({\bf{3}}, {\bf{2}})_{1/3,0,0}$ & $1/6$ & $q$ 
& $\eta \rho_1 \rho_2$ &  $Q^c$ & $\rho \overline{\eta}~\overline{\rho}_1 \overline{\rho}_2$ \\ \hline
$({\bf{3}}, {\bf{1}})_{-2/3,0,0}$ & $-1/3$ & $D$ 
& $\eta \rho_1 \rho_3$ &  $d^c$ & $\rho \overline{\eta}~\overline{\rho}_1 \overline{\rho}_3$ \\ \hline
$({\bf{1}}, {\bf{2}})_{-1/3, 1, 1/3}$ & $1/2$ & $L^c$ 
& $\eta \rho_2 \rho_3 \rho_4$ & $l$ 
& $\rho \overline{\eta}~\overline{\rho}_2 \overline{\rho}_3 \overline{\rho}_4$ \\ \hline
$({\bf{1}}, {\bf{2}})_{-1/3, -1, 1/3}$ & $-1/2$ & $l_1$ 
& $\eta \rho_2 \rho_3 \rho_5$ &  $L^c_1$  
& $\rho \overline{\eta}~\overline{\rho}_2 \overline{\rho}_3 \overline{\rho}_5$ \\ \hline
$({\bf{1}}, {\bf{2}})_{-1/3, 0, -2/3}$ & $-1/2$ & $l_2$ 
& $\eta \rho_2 \rho_3 \rho_6$ &  $L^c_2$ 
& $\rho \overline{\eta}~\overline{\rho}_2 \overline{\rho}_3 \overline{\rho}_6$ \\ \hline
$({\bf{1}}, {\bf{1}})_{2/3, 1, 1/3}$ & $1$ & $e^c$ 
& $\eta (\rho_2)^2 \rho_4$ &  $E$ 
& $\rho \overline{\eta} (\overline{\rho}_2)^2 \overline{\rho}_4$ \\ \hline
$({\bf{1}}, {\bf{1}})_{2/3, -1, 1/3}$ & $0$ & $\nu^c_1$ 
& $\eta (\rho_2)^2 \rho_5$ &  $N_1$ 
& $\rho \overline{\eta} (\overline{\rho}_2)^2 \overline{\rho}_5$ \\ \hline
$({\bf{1}}, {\bf{1}})_{2/3, 0, -2/3}$ & $0$ & $\nu^c_2$ 
& $\eta (\rho_2)^2 \rho_6$ &  $N_2$  
& $\rho \overline{\eta} (\overline{\rho}_2)^2 \overline{\rho}_6$ \\ \hline
$(\overline{\bf{3}}, {\bf{1}})_{0,-1,-1/3}$ & $-2/3$ & $u^c$ 
& $\eta (\rho_1)^2 \rho_5 \rho_6$ &  $U$  
& $\rho \overline{\eta} (\overline{\rho}_1)^2 \overline{\rho}_5 \overline{\rho}_6$ \\ \hline
$(\overline{\bf{3}}, {\bf{1}})_{0,1,-1/3}$ & $1/3$ & $d^c_1$ 
& $\eta (\rho_1)^2 \rho_4 \rho_6$ &  $D_1$  
& $\rho \overline{\eta} (\overline{\rho}_1)^2 \overline{\rho}_4 \overline{\rho}_6$ \\ \hline
$(\overline{\bf{3}}, {\bf{1}})_{0,0,2/3}$ & $1/3$ & $d^c_2$ 
& $\eta (\rho_1)^2 \rho_4 \rho_5$ &  $D_2$  
& $\rho \overline{\eta} (\overline{\rho}_1)^2 \overline{\rho}_4 \overline{\rho}_5$ \\ \hline
\end{tabular}
\end{center}
\end{table}
In the 4-th and 6-th column,
$\mathcal{P}_{f_{L}}$ and $\mathcal{P}_{(f_{R})^c}$ are the $Z_M$ element
of left-handed fermion and its charge conjugation of right-handed fermion, respectively.
The $\mathcal{P}_{(f_{R})^c}$ is determined by using (\ref{rhoPsi1R}).
The $\eta$ is the intrinsic $Z_M$ element of $\bf{27}$,
and $\overline{\rho}_i$ is the complex conjugation of $\rho_i$.
This assignment is applicable to the case with the extra space $S^1/Z_2$.

In Table \ref{t4}, the species and $Z_M$ element $\mathcal{P}_{b_{\alpha}}$ and $\mathcal{P}_{\overline{b}_{\alpha}}$ 
for multiplets from $({\bf{3}}, \overline{\bf{3}}, \overline{\bf{3}})$
and $(\overline{\bf{3}}, {\bf{3}}, {\bf{3}})$ of $G_{\tiny{\mbox{Tri}}}$ 
are given.
Here, their $Z_M$ element is determined by use of results for $SU(3)$ in Subsec. 2.3,
without considering that they are originated from ${\bf{78}}$ of $E_6$.
\begin{table}[htb]
\caption{The species and $Z_M$ element for gauge bosons.}
\label{t4}
\begin{center}
\begin{tabular}{l|c|c||l|c|c} \hline
\it{Representations} & $b_{\alpha}$ & $\mathcal{P}_{b_{\alpha}}$ & \it{Representations} 
& $\overline{b}_{\alpha}$ & $\mathcal{P}_{\overline{b}_{\alpha}}$ \\ \hline\hline
$({\bf{3}}, {\bf{2}})_{-1/3, -1, -1/3}$ & $b_{1}$ & $\rho_1 \rho_2 \rho_3 \rho_5 \rho_6$ & 
$(\overline{\bf{3}}, {\bf{2}})_{1/3,1,1/3}$ & $\overline{b}_1$ & $(\rho_1)^2 \rho_2 \rho_4$ \\ \hline
$({\bf{3}}, {\bf{2}})_{-1/3, 1, -1/3}$ & $b_{2}$ & $\rho_1 \rho_2 \rho_3 \rho_4 \rho_6$ & 
$(\overline{\bf{3}}, {\bf{2}})_{1/3,-1,1/3}$ & $\overline{b}_2$ & $(\rho_1)^2 \rho_2 \rho_5$ \\ \hline
$({\bf{3}}, {\bf{2}})_{-1/3, 0, 2/3}$ & $b_{3}$ & $\rho_1 \rho_2 \rho_3 \rho_4 \rho_5$ & 
$(\overline{\bf{3}}, {\bf{2}})_{1/3,0,-2/3}$ & $\overline{b}_3$ & $(\rho_1)^2 \rho_2 \rho_6$ \\ \hline
$({\bf{3}}, {\bf{1}})_{2/3, -1, -1/3}$ & $b_4$ & $\rho_1 (\rho_2)^2 \rho_5 \rho_6$ & 
$(\overline{\bf{3}}, {\bf{1}})_{-2/3,1,1/3}$ & $\overline{b}_4$ &  $(\rho_1)^2 \rho_3 \rho_4$ \\ \hline
$({\bf{3}}, {\bf{1}})_{2/3, 1, -1/3}$ & ${b}_5$ & $\rho_1 (\rho_2)^2 \rho_4 \rho_6$ & 
$(\overline{\bf{3}}, {\bf{1}})_{-2/3,-1,1/3}$ & $\overline{b}_{5}$ & $(\rho_1)^2 \rho_3 \rho_5$ \\ \hline
$({\bf{3}}, {\bf{1}})_{2/3, 0, 2/3}$ & $b_6$ & $\rho_1 (\rho_2)^2 \rho_4 \rho_5$ & 
$(\overline{\bf{3}}, {\bf{1}})_{-2/3,0,-2/3}$ & $\overline{b}_{6}$ & $(\rho_1)^2 \rho_3 \rho_6$ \\ \hline
\end{tabular}
\end{center}
\end{table}

{}From the $Z_M$ invariance of gauge kinetic term, the following relations are derived
\begin{eqnarray}
(\rho_1)^3 (\rho_2)^2 \rho_3 \rho_4 \rho_5 \rho_6 =1~,~~
(\rho_2)^2 \rho_3 \rho_4 \rho_5 \rho_6 =1~.
\label{R1}
\end{eqnarray}
The first relation comes from that the $Z_M$ element for the gauge boson
with complex-conjugate representation $\overline{\bm{R}}$ is the complex conjugation for that with $\bm{R}$.
In fact, the relation $\rho_1 \rho_2 \rho_3 \rho_5 \rho_6
=(\overline{\rho}_1)^2 \overline{\rho}_2 \overline{\rho}_4$ is required
for the $Z_M$ element of the pair $({\bf{3}}, {\bf{2}})_{-1/3, -1, -1/3}$
and $(\overline{\bf{3}}, {\bf{2}})_{1/3,1,1/3}$,
and the same relation is obtained for other pairs.
The second one comes from that all terms in the field strength $F_{MN}^a$
should possess a same $Z_M$ charge.
In fact, the product of $Z_M$ elements $({\bf{3}}, {\bf{2}})_{-1/3, -1, -1/3}$
and $({\bf{3}}, {\bf{2}})_{-1/3, 1, -1/3}$
should equal to that of $(\overline{\bf{3}}, {\bf{1}})_{-2/3,0,-2/3}$,
and the same relation is obtained for others.

{}From the $Z_M$ invariance of $\bf{27} \times \bf{27} \times \bf{27}$ up to an overall factor,\footnote{
We can construct $Z_M$ invariant terms such as 
$\bf{27}_a \times \bf{27}_b \times \bf{27}_c$ 
or $\bf{1} \times \cdots \times \bf{1} \times
\bf{27} \times \bf{27} \times \bf{27}$
by introducing differnt multiples and/or gauge singlets $\bf{1}$ 
with a suitable intrinsic $Z_M$ element.
}
we obtain the relation
\begin{eqnarray}
(\rho_1)^3 = (\rho_2)^2 \rho_3~,~~ \rho_4 \rho_5 \rho_6=1~.
\label{R2}
\end{eqnarray}

Combining (\ref{R1}) and (\ref{R2}), we obtain the relation
\begin{eqnarray}
(\rho_1)^3 =(\rho_2)^2 \rho_3 = \rho_4 \rho_5 \rho_6=1~,
\label{R2-2}
\end{eqnarray}
and $(\rho_1)^3 =1$ leads to 
\begin{eqnarray}
\rho_1=1~~~(M=2,4)~,~~~\rho_1=1, \omega, \omega^2~~~(M=3,6)~,
\label{R1-2}
\end{eqnarray}
where $\omega = e^{2\pi i/3}$.
It is shown that $\mathcal{P}_{b_{\alpha}}$ and $\mathcal{P}_{\overline{b}_{\alpha}}$
agree with those obtained from $\bf{27} \times \overline{\bf{27}}$ by use of (\ref{R2-2}).

Then we find that the $Z_M$ element for $(f_{R})^c$ from $\overline{\bf{27}}$
agrees with that for $f_{L}$ from $\bf{27}$
and the $Z_M$ element for $f_{L}$ from $\overline{\bf{27}}$
agrees with as that for $(f_{R})^c$ from $\bf{27}$,
if the intrinsic $Z_M$ element of $\overline{\bf{27}}$ is assigned by 
$\eta_{\overline{\bf{27}}}=\rho \overline{\eta}$.
Hence we study the classification based on $\bf{27}$.

Now let us impose the following conditions from a phenomenological point of view,
in order to reduce the assignment of $Z_M$ element.\\
1. The species $q$, $e^c$ and $u^c$ in each family of the SM survive 
as zero modes for each ${\bf 27}$ or $\overline{\bf 27}$, after compactification.\footnote{
There is a proposal that a large flavor mixing in lepton sector 
and a milder mass hierarchy of leptons and down-type quarks than up-type quarks
can be explained from a difference of origin for species.
That is, the species in ${\bf 10}$s of $SU(5)$ come from the corresponding ${\bf 27}$s, and
those in $\overline{\bf 5}$s come from the first two ${\bf 27}$s.\cite{B&M}
This interesting possibility would be excluded if we impose a stronger condition 
that all members of one family survive as zero modes after compactification.}\\
2. Most zero modes of mirror fermions (except SM singlets) are projected out by orbifolding.

Using the assignment of $Z_M$ element in Table \ref{t2} and the first condition, 
the relations $\eta \rho_1 \rho_2 = 1$, $\eta (\rho_2)^2 \rho_4 = 1$,
and $\eta (\rho_1)^2 \rho_5 \rho_6 = 1$ are required.

Combining them with (\ref{R1}) and (\ref{R2}), we derive the relations
\begin{eqnarray}
\rho_2 = \overline{\eta}~\overline{\rho}_1 = \overline{\eta} (\rho_1)^2~,~~ 
\rho_3 = \eta^2 \overline{\rho}_1 = \eta^2 (\rho_1)^2~,~~
\rho_4 = \eta (\rho_1)^2~,~~
\rho_6 = \overline{\eta} {\rho}_1 \overline{\rho}_5~.
\label{R3}
\end{eqnarray}
Hence the representation matrix is given by
\begin{eqnarray}
&~& R_M = \mbox{diag}(\rho_1, \rho_1, \rho_1) 
\times  \mbox{diag}(\overline{\eta} (\rho_1)^2, \overline{\eta} (\rho_1)^2, \eta^2 (\rho_1)^2)
\nonumber \\
&~& ~~~~~~~~~~~~~~~~ \times  \mbox{diag}(\eta (\rho_1)^2, \rho_5, \overline{\eta} {\rho}_1 \overline{\rho}_5)~,
\label{R-Y(1)-27}
\end{eqnarray}
where $\rho_1=1$ for $M=2, 4$ and $\rho_1=1, \omega, \omega^2$ for $M=3,6$.

Using (\ref{R-Y(1)-27}), the $Z_M$ element for $f_L$ is given by
\begin{eqnarray}
&~& \mathcal{P}_{q} = \mathcal{P}_{e^c} = \mathcal{P}_{u^c} = 1~,~~
\mathcal{P}_{D} = \mathcal{P}_{L^c} = \eta^3~,~~
\mathcal{P}_{l_1} = \eta^2 {\rho}_1 \rho_5~,
\nonumber \\
&~& \mathcal{P}_{l_2} = \eta (\rho_1)^2 \overline{\rho}_5~,~~
\mathcal{P}_{\nu^c_1} = \overline{\eta} {\rho}_1 \rho_5~,~~
\mathcal{P}_{\nu^c_2} = \overline{\eta}^2 (\rho_1)^2 \overline{\rho}_5~,
\nonumber \\
&~& \mathcal{P}_{d^c_1} = \eta ({\rho}_1)^2 \overline{\rho}_5~,~~
\mathcal{P}_{d^c_2} = \eta^2 {\rho}_1 {\rho}_5~.
\label{rhofL}
\end{eqnarray}
Then the following relations hold
\begin{eqnarray}
\mathcal{P}_{l_1} = \eta^3 \mathcal{P}_{\nu^c_1} = \eta^3 \overline{\mathcal{P}}_{d^c_1}~,~~
\eta^3 \overline{\mathcal{P}}_{l_2} = \overline{\mathcal{P}}_{\nu^c_2} = \mathcal{P}_{d^c_2}~.
\label{rhofL-rel}
\end{eqnarray}

In the same way, the $Z_M$ element for $(f_R)^c$ is given by
\begin{eqnarray}
&~& \mathcal{P}_{Q^c} = \mathcal{P}_{E} = \mathcal{P}_{U} = \rho~,~~
\mathcal{P}_{d^c} = \mathcal{P}_{l} = \rho \overline{\eta}^3~,~~
\mathcal{P}_{L^c_1} = \rho \overline{\eta}^2 ({\rho}_1)^2 \overline{\rho}_5~,
\nonumber \\
&~& \mathcal{P}_{L^c_2} = \rho \overline{\eta} {\rho}_1 {\rho}_5~,~~
\mathcal{P}_{N_1} = \rho {\eta} ({\rho}_1)^2 \overline{\rho}_5~,~~
\mathcal{P}_{N_2} = \rho {\eta}^2 {\rho}_1 {\rho}_5~,
\nonumber \\
&~& \mathcal{P}_{D_1} = \rho \overline{\eta} {\rho}_1 {\rho}_5~,~~
\mathcal{P}_{D_2} = \rho \overline{\eta}^2 (\rho_1)^2 \overline{\rho}_5~.
\label{rhofRc}
\end{eqnarray}
Then the following relations hold
\begin{eqnarray}
\mathcal{P}_{L^c_1} = \overline{\eta}^3 \mathcal{P}_{N_1} = \rho^2 \overline{\eta}^3 \overline{\mathcal{P}}_{D_1}~,~~
\rho^2 \overline{\eta}^3 \overline{\mathcal{P}}_{L^c_2} = \rho^2 \overline{\mathcal{P}}_{N_2} = \mathcal{P}_{D_2}~.
\label{rhofRc-rel}
\end{eqnarray}

For reference, the $Z_M$ element for extra gauge bosons with non-vanishing gauge quantum numbers is given by
\begin{eqnarray}
&~& \mathcal{P}_{b_1} = 1~,~~ \mathcal{P}_{b_2} = \eta ({\rho}_1)^2 \overline{\rho}_5~,~~
\mathcal{P}_{b_3} = \eta^2 {\rho}_1 \rho_5~,~~ \mathcal{P}_{b_4} = \overline{\eta}^3~,
\nonumber \\
&~& \mathcal{P}_{b_5} = \overline{\eta}^2 (\rho_1)^2 \overline{\rho}_5~,~~
\mathcal{P}_{b_6} = \overline{\eta} {\rho}_1 \rho_5~,~~
\mathcal{P}_{b_7} = \overline{\eta}^3~,~~
\mathcal{P}_{b_8} = \eta ({\rho}_1)^2 \overline{\rho}_5~,
\nonumber \\
&~& \mathcal{P}_{b_9} = \eta ({\rho}_1)^2 ({\rho}_5)^2~,~~
\mathcal{P}_{b_{10}} = \overline{\eta}^2 ({\rho}_1)^2 \overline{\rho}_5~,
\label{Pbalpha}
\end{eqnarray}
where $b_7$ is $({\bf{1}}, {\bf{2}})_{1,0,0}$, and
$b_8$, $b_9$ and $b_{10}$ are 
$({\bf{1}}, {\bf{1}})_{0,2,0}$, $({\bf{1}}, {\bf{1}})_{0,-1,1}$, and $({\bf{1}}, {\bf{1}})_{0,-1,-1}$, respectively.
The $Z_M$ element for $\overline{b}_{\alpha}$ is given by
$\mathcal{P}_{\overline{b}_{\alpha}} = \overline{\mathcal{P}}_{b_{\alpha}}$.

Here, we point out generic features.\\
(a) The mirror fermions such as $Q^c$, $E$ and $U$ are always projected out,
because they have the $Z_M$ element $\rho = e^{2\pi i/M} \ne 1$.\\
(b) The mirror fermions such as $D$ and $L^c$ have a same $Z_M$ element
and are projected out, if $d^c$ and $l$ have zero mode.\\
(c) The extra gauge bosons with $({\bf{3}}, {\bf{2}})_{-1/3, -1, -1/3}$ and  
$(\overline{\bf{3}}, {\bf{2}})_{1/3,1,1/3}$ always have zero modes,
and $E_6$ does not break down to $G_{\tiny{\mbox{Tri}}}$ and its subgroups
in our setup.

Because the representation matrix based on $Y_{(2)}$ ($Y_{(3)}$)
is obtained by exchanging $\rho_4$ and $\rho_5$ ($\rho_4$ and $\rho_6$),
we obtain same results irrespective of the definition of hypercharge.
Hence we consider the case with $Y_{(1)}$, in the following.

\subsection{$M=2$}

The representation matrix for $M=2$ is given by
\begin{eqnarray}
R_{2} = \mbox{diag}(1, 1, 1) \times
\mbox{diag}(\eta, {\eta}, 1) \times
\mbox{diag}(\eta, \rho_5, {\eta} {\rho}_5)~,
\label{R-M=2}
\end{eqnarray}
where we use $\overline{\eta} = \eta$, $\rho_1=1$ and $\overline{\rho}_5 = \rho_5$, .

The matrix (\ref{R-M=2}) is characterized by $\eta$ and $\rho_5$.
The $Z_2$ element and species with even parity are given in Table \ref{t5}.
\begin{table}[htb]
\caption{The $Z_2$ element and species with even parity from $\bf{27}$.}
\label{t5}
\begin{center}
\begin{tabular}{c|c||l|c} \hline
$\eta$ & $\rho_5$  & \it{Species with even parity} & \it{Gauge group}
\\ \hline\hline
$1$ & $1$ & 
$q$, $e^c$, $u^c$, $D$, $L^c$, $l_1$, $l_2$, $d^c_1$, $d^c_2$, $\nu^c_1$, $\nu^c_2$& $E_6$\\ \hline
$1$ & $-1$ & 
$q$, $e^c$, $u^c$, $D$, $L^c$, $[L^c_1]$, $[L^c_2]$, $[D_1]$, $[D_2]$, $[N_1]$, $[N_2]$& $SU(6)\times SU(2)$ \\ \hline
$-1$ & $1$ & 
$q$, $e^c$, $u^c$, $[d^c]$, $[l]$, $l_1$, $[L^c_2]$, $[D_1]$, $d^c_2$, $[N_1]$, $\nu^c_2$& $SO(10)\times U(1)$ \\ \hline
$-1$ & $-1$ & 
$q$, $e^c$, $u^c$, $[d^c]$, $[l]$, $[L^c_1]$, $l_2$, $d^c_1$, $[D_2]$, $\nu^c_1$, $[N_2]$& $SO(10)\times U(1)$ \\ \hline
\end{tabular}
\end{center}
\end{table}
In Table \ref{t5}, the species in square bracket originate from the charge conjugation of
right-handed one.
For reference, we list also a case that $d^c$ and $l$ are absent.

For reference, we give a correspondence between the shift embedding on roots of $E_6$ and ours.
The second one in Table \ref{t5} is realized by the shift vector $\overline{V}=(0,1/2,1/2,0,0,1/2)$,
and the third and fourth ones are realized by $\overline{V}=(1/2,1/2,0,1/2,1/2,0)$.
The difference of third and fourth ones stems from that of assignment of species.
Here, we use the gauge shift $\overline{V}$ defined in \cite{BK&R}.

Zero modes are reduced by the combination of representation matrices,
and they are given by the intersection of them.
Here, we give an example with all members of one SM family (except a neutrino singlet) 
in the smallest gauge group.
For $S^1/Z_2$, the combination of representation matrices such that
\begin{eqnarray}
&~& P_{0} = \mbox{diag}(1, 1, 1) \times \mbox{diag}(-1, -1, 1) \times \mbox{diag}(-1, 1, -1)
\label{R-M=21}\\
&~& P_{1} = \mbox{diag}(1, 1, 1) \times \mbox{diag}(-1, -1, 1) \times \mbox{diag}(-1, -1, 1)
\label{R-M=22}
\end{eqnarray}
generates zero modes of $(q, e^c, u^c, [d^c], [l])$.
The gauge group is $SU(5) \times U(1)^2$.
For $T^2/Z_2$, the same matters and gauge group are obtained with $P_2 = P_0$ or $P_1$, 
as well as the above $P_0$ and $P_1$.

\subsection{$M=3$}

The representation matrix for $M=3$ is given by
\begin{eqnarray}
&~& R_{3} = \mbox{diag}(\rho_1, \rho_1, \rho_1) \times
\mbox{diag}(\eta^2 ({\rho}_1)^2, {\eta}^2 ({\rho}_1)^2, \eta^2 ({\rho}_1)^2) 
\nonumber \\
&~& ~~~~~~~~~~~~~~~ \times \mbox{diag}(\eta (\rho_1)^2, \rho_5, {\eta}^2 \rho_1 ({\rho}_5)^2)~,
\label{R-M=3}
\end{eqnarray}
where we use $\overline{\eta} = \eta^2$, $\overline{\rho}_1 = (\rho_1)^2$ and $\overline{\rho}_5 = (\rho_5)^2$.
The following relations hold
\begin{eqnarray}
&~& \mathcal{P}_{q} = \mathcal{P}_{e^c} = \mathcal{P}_{u^c} = \mathcal{P}_{D} = \mathcal{P}_{L^c} = 1~,
\nonumber \\
&~& \mathcal{P}_{l_1} = \mathcal{P}_{\nu^c_1} = \overline{\mathcal{P}}_{d^c_1} =
\overline{\mathcal{P}}_{l_2} = \overline{\mathcal{P}}_{\nu^c_2} = \mathcal{P}_{d^c_2} = \eta^2 \rho_1 \rho_5~,~~
\label{rhofL-rel-M=3}\\
&~& \mathcal{P}_{Q^c} = \mathcal{P}_{E} = \mathcal{P}_{U} = \mathcal{P}_{d^c} = \mathcal{P}_{l} = \rho~,
\nonumber \\
&~& \mathcal{P}_{L^c_1} = \mathcal{P}_{N_1} = \rho^2 \overline{\mathcal{P}}_{D_1} 
= \rho^2 \overline{\mathcal{P}}_{L^c_2} = \rho^2 \overline{\mathcal{P}}_{N_2} 
= \mathcal{P}_{D_2} = \rho \eta ({\rho}_1)^2 (\rho_5)^2~.
\label{rhofRc-rel-M=3}
\end{eqnarray}

In this way, zero modes of mirror particles such as $D$ and $L^c$ always appear.
Zero modes of ${l_1}$, ${l_2}$, ${\nu^c_1}$, ${\nu^c_2}$, ${d^c_1}$
and ${d^c_2}$ can survive together in the case with $\rho_1 \rho_5=\eta$.
Then the mirror particles such as $L^c_a$ and $D_a$ are projected out because of
$\mathcal{P}_{L^c_1} = \mathcal{P}_{N_1} = \rho^2 \overline{\mathcal{P}}_{D_1} 
= \rho^2 \overline{\mathcal{P}}_{L^c_2} = \rho^2 \overline{\mathcal{P}}_{N_2} 
= \mathcal{P}_{D_2} = \rho$.
Hence the fermions such as $q$, $e^c$, $u^c$, $l_a$, $\nu^c_a$, $d^c_a$, $D$ and $L^c$ $(a=1,2)$
survive after compactification and $E_6$ is unbroken, in the case with $\rho_1 \rho_5=\eta$.

\subsection{$M=4$}

The representation matrix for $M=4$ is given by
\begin{eqnarray}
R_{4} = \mbox{diag}(1, 1, 1) \times
\mbox{diag}(\eta^3, {\eta}^3, \eta^2) 
\times \mbox{diag}(\eta, \rho_5, {\eta}^3 ({\rho}_5)^3)~,
\label{R-M=4}
\end{eqnarray}
where we use $\overline{\eta} = \eta^3$, $\rho_1=1$ and $\overline{\rho}_5 = (\rho_5)^3$.
The following relations hold
\begin{eqnarray}
\hspace{-1cm}&~& \mathcal{P}_{q} = \mathcal{P}_{e^c} = \mathcal{P}_{u^c} = 1~,~~
\mathcal{P}_{D} = \mathcal{P}_{L^c} = \eta^3~,
\nonumber \\
\hspace{-1cm}&~& \mathcal{P}_{l_1} = \eta^3 \mathcal{P}_{\nu^c_1} = \eta^3 \overline{\mathcal{P}}_{d^c_1} 
= \eta^2 \rho_5~,~~
\nonumber \\
\hspace{-1cm}&~& \eta^3 \overline{\mathcal{P}}_{l_2} = \overline{\mathcal{P}}_{\nu^c_2} 
= \mathcal{P}_{d^c_2} = \eta^2 \rho_5~,~~
\label{rhofL-rel-M=4}\\
\hspace{-1cm}&~& \mathcal{P}_{Q^c} = \mathcal{P}_{E} = \mathcal{P}_{U} = \rho~,~~
\mathcal{P}_{d^c} = \mathcal{P}_{l} = \rho \eta~,
\nonumber \\
\hspace{-1cm}&~& \mathcal{P}_{L^c_1} = \eta \mathcal{P}_{N_1} = \rho^2 \eta \overline{\mathcal{P}}_{D_1} 
= \rho \eta^2 (\rho_5)^3~,~~
\nonumber \\
\hspace{-1cm}&~& \rho^2 \eta \overline{\mathcal{P}}_{L^c_2} = \rho^2 \overline{\mathcal{P}}_{N_2} 
= \mathcal{P}_{D_2} = \rho \eta^2 (\rho_5)^3~.
\label{rhofRc-rel-M=4}
\end{eqnarray}

Here, we consider a possibility that all charged mirror fermions are projected out 
by $Z_4$ orbifolding, for simplicity.
The value of intrinsic $Z_4$ element is determined as $\eta =i, -1, -i$ to project out $D$ and $L^c$.
The value of $\rho_5$ is determined as $\rho_5 = i, -1$ for $\eta = i$,
$\rho_5 = 1, -1, -i$ for $\eta = -1$ and $\rho_5 = 1, i$ for $\eta = -i$ to project out $D_2$ and $L^c_2$.

The $Z_4$ element and species with zero modes are given in Table \ref{t6}.
The species in square bracket originate from the charge conjugation of
right-handed one.
For reference, we list also a case that $d^c$ and $l$ are absent.
\begin{table}[htb]
\caption{The $Z_4$ element and species with zero modes from $\bf{27}$.}
\label{t6}
\begin{center}
\begin{tabular}{c|c||l||c} \hline
$\eta$ & $\rho_5$  & \it{Species with zero modes} & \it{Gauge group}
\\ \hline\hline
$i$ & $i$& $q$, $e^c$, $u^c$, $l_2$, $d^c_1$, $\nu^c_1$, $[N_2]$ & $SO(10)\times U(1)$ \\ \hline
$i$ & $-1$& $q$, $e^c$, $u^c$, $l_1$, $d^c_2$, $\nu^c_2$, $[N_1]$ & $SO(10)\times U(1)$ \\ \hline 
$-1$ & $-i$ & $q$, $e^c$, $u^c$, $[N_1]$, $[N_2]$ & $SU(5)\times SU(2)\times U(1)$  \\ \hline
$-1$ & 1 & $q$, $e^c$, $u^c$, $l_1$, $d^c_2$, $\nu^c_2$ & $SO(10)\times U(1)$  \\ \hline
$-1$ & $-1$& $q$, $e^c$, $u^c$, $l_2$, $d^c_1$, $\nu^c_1$ & $SO(10)\times U(1)$ \\ \hline
$-i$ & 1& $q$, $e^c$, $u^c$, $[d^c]$, $[l]$,  $[N_1]$ & $SU(5)\times U(1)^2$ \\ \hline
$-i$ & $i$& $q$, $e^c$, $u^c$, $[d^c]$, $[l]$, $[N_2]$ & $SU(5)\times U(1)^2$ \\ \hline
\end{tabular}
\end{center}
\end{table}

Here, we give an example with all members of one SM family (except a neutrino singlet) 
in the smallest gauge group.
The combination of representation matrix for $Z_4$ and $Z_2$ such as
\begin{eqnarray}
&~& Q_{0} = \mbox{diag}(1, 1, 1) \times \mbox{diag}(i, i, -1) \times \mbox{diag}(-i, 1, i)~,
\label{R-M=44}\\
&~& P_{1} = \mbox{diag}(1, 1, 1) \times \mbox{diag}(1, 1, 1) \times \mbox{diag}(1, 1, 1)~,
\label{R-M=42}
\end{eqnarray}
we have a model with just SM family members
and a gauge singlet, i.e., zero modes of $q$, $e^c$, $u^c$, $[d^c]$, $[l]$ and $[N_1]$,
and the gauge group $SU(5) \times U(1)^2$.

\subsection{$M=6$}

The representation matrix for $M=6$ is given by
\begin{eqnarray}
&~& R_{6} = \mbox{diag}(\rho_1, \rho_1, \rho_1) \times
\mbox{diag}(\eta^5 ({\rho}_1)^2, {\eta}^5 ({\rho}_1)^2, \eta^2 ({\rho}_1)^2) 
\nonumber \\
&~& ~~~~~~~~~~~~~~~ \times \mbox{diag}(\eta (\rho_1)^2, \rho_5, {\eta}^5 \rho_1 ({\rho}_5)^5)~,
\label{R-M=6}
\end{eqnarray}
where we use $\overline{\eta} = \eta^5$, $(\rho_1)^3 = 1$ and $\overline{\rho}_5 = (\rho_5)^5$.
The following relations hold
\begin{eqnarray}
\hspace{-1.4cm}&~& \mathcal{P}_{q} = \mathcal{P}_{e^c} = \mathcal{P}_{u^c} = 1~,~~
\mathcal{P}_{D} = \mathcal{P}_{L^c} = \eta^3~,
\nonumber \\
\hspace{-1.4cm}&~& \mathcal{P}_{l_1} = \eta^3 \mathcal{P}_{\nu^c_1} = \eta^3 \overline{\mathcal{P}}_{d^c_1}
= \eta^3 \overline{\mathcal{P}}_{l_2} = \overline{\mathcal{P}}_{\nu^c_2} 
= \mathcal{P}_{d^c_2} = \eta^2 \rho_1 \rho_5~,
\label{rhofL-rel-M=6}\\
\hspace{-1.4cm}&~& \mathcal{P}_{Q^c} = \mathcal{P}_{E} = \mathcal{P}_{U} = \rho~,~~
\mathcal{P}_{d^c} = \mathcal{P}_{l} = \rho \eta^3~,
\nonumber \\
\hspace{-1.4cm}&~& \mathcal{P}_{L^c_1} = \eta^3 \mathcal{P}_{N_1} = \rho^2 \eta^3 \overline{\mathcal{P}}_{D_1} 
= \rho^2 \eta^3 \overline{\mathcal{P}}_{L^c_2} = \rho^2 \overline{\mathcal{P}}_{N_2} 
= \mathcal{P}_{D_2} = \rho \eta^4 ({\rho}_1)^2 (\rho_5)^5~.
\label{rhofRc-rel-M=6}
\end{eqnarray}
Note that $d^c$ and $l$ are always projected out.

The assignment of $Z_6$ element and gauge group are given in Table \ref{t6-2}.
\begin{table}[htb]
\caption{The assignment of $Z_6$ element for $\bf{27}$.}
\label{t6-2}
\begin{center}
\begin{tabular}{c|c|c} \hline
& $(\eta,\rho_1)$ & \it{Gauge group}  
\\ \hline\hline
$\eta^2 \rho_1 \rho_5=1$ & $(\rho^m,\rho^n)~(m=1,3,5,n=0,2,4)$ & $SO(10)\times U(1)$ \\ \hline
$\eta^5 \rho_1 \rho_5=1$ & $(\rho^m,\rho^n)~(m=1,3,5,n=0,2,4)$ & $SO(10)\times U(1)$  \\ \hline
\end{tabular}
\end{center}
\end{table}
Here, we consider a possibility that all charged mirror fermions are projected out 
and kept all members of one family ($q$, $e^c$,$u^c$, $l_1$, $\nu^c_2$, $d^c_2$)
or ($q$, $e^c$,$u^c$, $l_2$, $\nu^c_1$, $d^c_1$) by $Z_6$ orbifolding, for simplicity.
The value of intrinsic $Z_6$ element is determined as $\eta =\rho, \rho^3, \rho^5$ $(\rho \equiv e^{\pi i/3})$
to project out $D$ and $L^c$.
If we choose $\eta^2 \rho_1 \rho_5 =1$ to survive zero modes of $l_1$, $\nu^c_2$ and $d^c_2$,
zero modes of other fermions except for $q$, $e^c$ and $u^c$ are projected out
because of 
$\mathcal{P}_{L^c_1} = \eta^3 \mathcal{P}_{N_1} = \rho^2 \eta^3 \overline{\mathcal{P}}_{D_1} 
= \rho^2 \eta^3 \overline{\mathcal{P}}_{L^c_2} = \rho^2 \overline{\mathcal{P}}_{N_2} 
= \mathcal{P}_{D_2} = \rho$.
In the same way, if we choose $\eta^5 \rho_1 \rho_5 =1$, 
zero modes of $q$, $e^c$,$u^c$, $l_2$, $\nu^c_1$ and $d^c_1$
survive and those of others are projected out.

By a suitable combination of representation matrix for $Z_6$ and $Z_2$, e.g.,
\begin{eqnarray}
&~& R_{6} = \mbox{diag}(1, 1, 1) \times \mbox{diag}(\rho^5, \rho^5, \rho^2) \times \mbox{diag}(\rho, \rho, \rho^4)~,
\label{R-M=66}\\
&~& R_{2} = \mbox{diag}(1, 1, 1) \times \mbox{diag}(1, 1, 1) \times \mbox{diag}(1, 1, 1)~,
\label{R-M=62}
\end{eqnarray}
we have a model with just SM family members
and a gauge singlet, i.e., zero modes of $q$, $e^c$,$u^c$, $l_2$, $\nu^c_1$ and $d^c_1$, 
and the unbroken gauge group $SO(10) \times U(1)$.

\section{Effective grand unified model}

The string-inspired $E_6$ SUSY grand unified theories have been studied intensively
since the construction of 4-dimensional models 
based on the Calabi-Yau compactification.\cite{CY}\footnote{
For the theoretical and phenomenological aspects
of string-inspired $E_6$ models, see \cite{KM&N} and references therein.}
$E_6$ grand unified theories with three generations have been derived from heterotic string theory.\cite{E6het}
Higher-dimensional $E_6$ grand unified theories on orbifold have been also studied
from several aspects.\cite{HJL&L,H&S,H&R,FN&W,BK&R}

Most low-energy theories derived from higher-dimensional $E_6$ grand unified theories
contain exotic particles such as $D$, $L^c$ and so on, in our notation.
They influence the gauge coupling unification and can induce the problem of proton decay.
Hence it is interesting to derive an effective grand unified model, which
contains a minimal set of particle contents 
(if possible, exotic particles are absent in its low-energy theory)
and to study features of the model.
Two conditions in subsection 3.1 have been imposed on $Z_M$ elements of matter fields from this point of view.

We take the grand unified model whose gauge group is $SU(5) \times U(1)^2$,
derived from the $Z_4$ orbifolding on $T^2$,
with the representation matrices such that
\begin{eqnarray}
&~& Q_{0} = \mbox{diag}(1, 1, 1) \times \mbox{diag}(i, i, -1) \times \mbox{diag}(-i, 1, i)~,
\label{R-M=44-1}\\
&~& P_{1} = \mbox{diag}(1, 1, 1) \times \mbox{diag}(1, 1, 1) \times \mbox{diag}(1, 1, 1)~.
\label{R-M=42-2}
\end{eqnarray}
This model has zero modes of $(q, e^c, u^c, [d^c], [l], [N_1])$,
which belong to the fields of $({\bf{10}}, -1, 1)$, $(\overline{\bf{5}}, -2, 2)$ and $({\bf{1}}, 0, -4)$
under $SU(5) \times U(1)_1 \times U(1)_2$.
Here, $SU(5) \times U(1)_1$ is a maximal subgroup of $SO(10)$,
and the normalization of $U(1)_1$ charge ($Q_1$) and $U(1)_2$ charge ($Q_2$)
are taken as $\sum_{\bf{27}} Q_1^2 = 120$ and $\sum_{\bf{27}} Q_2^2 = 72$, respectively.
The left-handed fermions of $(\overline{\bf{5}}, -2, 2)$ and $({\bf{1}}, 0, -4)$ are obtained 
by the charge-conjugation 
for right-handed fermion of $({\bf{5}}, 2, -2)$ and $({\bf{1}}, 0, 4)$, respectively.
The $[N_1]$s are regarded as neutrino singlets, which are involved the see-saw mechanism.\cite{see-saw1,see-saw2}
Note that there appear no exotic particles as zero modes from the bulk fields.

{}From the observation that the gauge couplings are unified 
in the minimal supersymmetric standard model (MSSM),
we assume that our model possesses SUSY, 
which is broken by the Scherk-Schwarz mechanism
in the bulk,\cite{S&S,S&S2}
and the gauge symmetry $SU(5) \times U(1)_1 \times U(1)_2$ 
is broken down to the SM one $G_{\tiny{\mbox{SM}}}$
at the unification scale $M_{\tiny{\mbox{GUT}}} (= 2.1 \times 10^{16})$GeV by the Higgs mechanism
due to localized fields on a fixed point.
Through the Scherk-Schwarz mechanism, bulk fields obtain a common soft SUSY breaking mass $m_0$.
On the other hand, localized fields are, in general, supposed to acquire 
non-universal soft SUSY breaking masses by other SUSY breaking sources.

First we give a prediction to test our model.
In 4-dimensional $E_6$ grand unified models, 
there are proposals that fermion masses can be useful probes\cite{BK&Y},
and sfermion masses can be also useful to know a pattern of gauge symmetry breaking
in the SUSY extensions.\cite{K&T,Kolda&Martin}\footnote{
Sfermion mass relations have been also studied in orbifold family unification models.\cite{K&K}
}
Hence it is interesting to study sum rules among superparticle masses 
such as sfermion masses and gaugino masses based on the SUSY extension of our model.

After the breakdown of $SU(5) \times U(1)_1 \times U(1)_2$, 
we have the following mass formulae at $M_{\tiny{\mbox{GUT}}}$,
\begin{eqnarray}
&~& m_{\bf{10}}^2 \equiv m_{\tilde{q}}^2 = m_{\tilde{u}^c}^2 = m_{\tilde{e}^c}^2 = m_0^2  -  D_1 + D_2~,
\label{m10}\\
&~& m_{\overline{\bf{5}}}^2 \equiv m_{\tilde{d}^c}^2 = m_{\tilde{l}}^2 = m_0^2 - 2 D_1 + 2 D_2~,
\label{mbar5}\\
&~& M_{\bf{24}} \equiv M_{3} = M_{2} = M_{1} = m_0~,~~M_{1_1} = M_{1_2} = m_0~,
\label{M24}
\end{eqnarray}
where $m_{\bf{10}}$ and $m_{\overline{\bf{5}}}$ are the soft SUSY breaking scalar masses for 
sfermions of $({\bf{10}}, -1, 1)$ and $(\overline{\bf{5}}, -2, 2)$, 
$D_1$ and $D_2$ are parameters which present $D$-term contributions relating $U(1)_1$ and $U(1)_2$,
and $M_{\bf{24}}$ is the soft SUSY breaking gaugino mass for the $SU(5)$ gaugino.
The $M_{3}$, $M_{2}$ and $M_{1}$ are gaugino masses of the SM group,
and $M_{1_1}$ and $M_{1_2}$ are gaugino masses of $U(1)_1$ and $U(1)_2$, respectively.

The $D$-term contributions, in general, originate from $D$-terms related to broken gauge symmetries 
when soft SUSY breaking parameters possess a non-universal structure in the gauge symmetry breaking sector 
and the rank of gauge group decreases after the breakdown of gauge symmetry.\cite{Dterm1,Dterm2,KMY1,KMY2}
In most cases, the magnitude of $D$-term condensation is, at most, of order original soft SUSY breaking mass squared,
and hence $D$-term contributions can induce sizable effects on sfermion spectrum.

By eliminating unknown parameters $D_1$ and $D_2$,
we obtain the specific relation
\begin{eqnarray}
2(m_{\bf{10}}^2 - M_{\bf{24}}^2) = m_{\overline{\bf{5}}}^2 - M_{\bf{24}}^2~.
\label{SR1}
\end{eqnarray}
Because the formulae (\ref{m10}) and (\ref{mbar5}) are generation-independent,
the following relations are also derived,
\begin{eqnarray}
m_{{\bf{10}}_1}^2 = m_{{\bf{10}}_2}^2 = m_{{\bf{10}}_3}^2~,~~
m_{\overline{\bf{5}}_1}^2 = m_{\overline{\bf{5}}_2}^2 = m_{\overline{\bf{5}}_3}^2~,
\label{SR2}
\end{eqnarray}
where $m_{{\bf{10}}_i}$ and $m_{\overline{\bf{5}}_i}$ $(i=1, 2, 3)$ 
are soft SUSY breaking scalar masses for the $i$-th generation.
The sum rules (\ref{SR1}) and (\ref{SR2}) can be useful probes for our model.

Next we discuss the structure of superpotential and problems relating it.
We consider a model with a minimal particle content, for simplicity.
Based on the extension of brane world scenario, our 4-dimensional world
is assumed to be the space-time fixed on the origin of $T^2/Z_4$.
On our space-time, $SU(5) \times U(1)_1 \times U(1)_2$ gauge symmetry
is respected on the compactification.

Let us introduce several chiral superfields on the fixed point of $Z_4$ transformation, i.e.,
 a chiral multiplet $\Sigma$ to break $SU(5)$ down to the SM one,
two pairs of chiral multiplets $(S_1, \overline{S}_1)$
and $(S_2, \overline{S}_2)$ to break $U(1)_1$ and $U(1)_2$,
and a pair of chiral multiplet $(H, \overline{H})$ 
to break the electroweak symmetry down to the electric one.
The gauge quantum numbers of such localized fields are given in Table \ref{t7}.
Other charged fields are necessary to cancel anomalies relating $U(1)_1$ and $U(1)_2$.
\begin{table}[htb]
\caption{The gauge quantum numbers of localized fields.}
\label{t7}
\begin{center}
\begin{tabular}{c||c|c||c} \hline
\it{Localized fields} & $SU(5)$  & $U(1)_1$ & $U(1)_2$
\\ \hline\hline
$\Sigma$ & $\bf{24}$ & $0$ & $0$ \\ \hline
$S_1$ & $\bf{1}$ & $1$ & $0$ \\ \hline
$\overline{S}_1$ & $\bf{1}$ & $-1$ & $0$ \\ \hline
$S_2$ & $\bf{1}$ & $0$ & $1$ \\ \hline
$\overline{S}_2$ & $\bf{1}$ & $0$ & $-1$ \\ \hline
$H$ & $\bf{5}$ & $0$ & $0$ \\ \hline
$\overline{H}$ & $\overline{\bf{5}}$ & $0$ & $0$ \\ \hline
\end{tabular}
\end{center}
\end{table}

The superpotential $W$ is given by
\begin{eqnarray}
&~& W = \frac{\tilde{f}_{U}^{ij}}{\Lambda^4} (S_1 \overline{S}_2)^2 {\bf{10}}_i {\bf{10}}_j H +
\frac{\tilde{f}_{D}^{ij}}{\Lambda^6} (S_1 \overline{S}_2)^3 {\bf{10}}_i \overline{\bf{5}}_j \overline{H} 
+ \frac{M^{ij}}{\Lambda^4} {S_2}^4 {\bf{1}}_i {\bf{1}}_j
\nonumber \\
&~& ~~~~~~~ + f_{\Sigma} \overline{H} \Sigma H + \mu_{H} \overline{H} H + W_S~,
\label{W}
\end{eqnarray}
where $\Lambda$ is a cutoff scale and
$W_S$ is a superpotential which induces the breakdown of $U(1)_1 \times U(1)_2$,
e.g., $W_S = f_S \tilde{S} (S_1 \overline{S}_1 - M^2) + f_{S'} \tilde{S}' (S_2 \overline{S}_2 - M^2) + \cdots$,
using $SU(5)$ singlet chiral multiplets $\tilde{S}$ and $\tilde{S}'$.
We impose $R$-parity invariance on $W$.
The vacuum expectation value (VEV) of scalar fields is determined from 
a minimum of scalar potential including soft SUSY breaking terms.

The scalar component of 
$\Sigma$ acquire the VEV of $\langle \Sigma \rangle = \mbox{diag}(2, 2, 2, -3, -3) V$,
and $SU(5)$ is broken down to $G_{\tiny{\mbox{SM}}}$.
The scalar components of $(S_1, \overline{S}_1)$ and $(S_2, \overline{S}_2)$ 
acquire the VEVs of order $M_{\tiny{\mbox{GUT}}}$,
and $U(1)_1 \times U(1)_2$ is broken down.
Then the superpotential becomes the effective one,
\begin{eqnarray}
W_{\tiny{\mbox{eff}}} = f_{U}^{ij} {\bf{10}}_i {\bf{10}}_j H +
f_{D}^{ij} {\bf{10}}_i \overline{\bf{5}}_j \overline{H} + \mu \overline{H}_W H_W
+ \mu_C \overline{H}_C H_C~,
\label{Weff}
\end{eqnarray}
where $H_W$ and $\overline{H}_W$ are weak Higgs doublets, 
$H_C$ and $\overline{H}_C$ are colored Higgs triplets,
and $f_{U}^{ij}$, $f_{D}^{ij}$, $\mu$ and $\mu_C$ are given by
\begin{eqnarray}
&~& f_{U}^{ij} = \frac{\tilde{f}_{U}^{ij}}{\Lambda^4} (\langle S_1 \rangle \langle \overline{S}_2 \rangle)^2~,~~
f_{D}^{ij} = \frac{\tilde{f}_{D}^{ij}}{\Lambda^6} (\langle S_1 \rangle \langle \overline{S}_2 \rangle)^3~,~~
\label{fij}\\
&~& \mu = - 3 f_{\Sigma} V + \mu_{H}~,~~\mu_C = 2 f_{\Sigma} V + \mu_{H}~.
\label{mu}
\end{eqnarray}

The $W_{\tiny{\mbox{eff}}}$ has a same form 
derived from the ordinary $SU(5)$ SUSY GUT,\cite{SUSYGUT1,SUSYGUT2}
and it has the fermion mass relation $m_{\tau} = m_b$ at $M_{\tiny{\mbox{GUT}}}$,
and induces problems relating the proton decay\cite{Proton1,Proton2} and the fine-tuning of Higgs masses.
It is future work to solve the problems by extending the minimal one
and to study other effective theories derived from $E_6$ orbifold grand unification.

\section{Conclusions and discussion}

We have classified the standard model particles, which originate from
bulk fields with $\bf{27}$ or $\overline{\bf{27}}$ after orbifold breaking,
in $E_6$ grand unified theories on 5 or 6-dimensional space-time,
and found that standard model family members survive 
under relatively big gauge groups such as $SO(10) \times U(1)$ and $SU(5) \times U(1)^2$ 
after orbifolding, based on the condition that $q$, $e^c$ and $u^c$ survive as zero modes
for each $\bf{27}$ or $\overline{\bf{27}}$.
We have studied features of SUSY $SU(5) \times U(1)_1 \times U(1)_2$ grand unified model
and found that sum rules among superparticle masses can be useful probes to test our model.

Our models can be a starting point to study a realistic grand unified theory.
There are several problems in the minimal version, which are left future work.
With the minimal particle contents, the $SU(5) \times U(1)^2$ grand unified models after orbifolding 
leads to a same type of superpotential of ordinary 4-dimensional SUSY $SU(5)$ GUT, 
and then it induces problems relating the proton decay and
the fine-tuning of Higgs masses.
It is interesting to solve the problems by extending the minimal one
and to study other effective theories derived from $E_6$ orbifold grand unification.

As another path, there is a possibility that 
smaller gauge groups such as $G_{\tiny{\mbox{Tri}}}$ and $G_{\tiny{\mbox{32111}}}$
are obtained directly through orbifolding,
if the condition on $q$, $e^c$ and $u^c$ is relaxed.
On behalf of it, extra bulk and/or localized fields should be introduced.

The Hosotani mechanism\cite{H1,H2} has been applied to the breakdown of unified gauge symmetry.\cite{KTY,Yamashita}
It is intriguing to construct models incorporating the Hosotani mechanism 
in the framework of $E_6$ grand unification.

Furthermore it is interesting to explore the origin of three families.
Orbifold family unification models may give us a hint.\cite{KK&O,K&M3}

\section*{Acknowledgements}
This work was supported in part by scientific grants from the Ministry of Education, Culture,
Sports, Science and Technology under Grant Nos.~22540272 and 21244036 (Y.K.)
and No.~23$\cdot$9368 (T.M.).

\end{document}